\documentclass[preprint,showpacs,pre,12pt]{revtex4-1}
\usepackage{amsfonts}
\usepackage{amssymb}
\usepackage{amsmath}
\usepackage{graphicx}

\begin{document}
\title{Perturbation approach to multifractal dimensions for certain critical random matrix ensembles}
\author{E. Bogomolny$^{1,2}$, O. Giraud$^{1,2}$}
\affiliation{$^{1}$Univ. Paris-Sud, 
Laboratoire de Physique Th\'eorique et Mod\`eles Statistiques, UMR8626, 91405 Orsay,
France\\
$^2$CNRS, 91405 Orsay, France}
\date{June 29, 2011}
\pacs{05.45.-a, 05.45.Df, 05.45.Mt, 71.30.+h}

\begin{abstract}
Fractal dimensions of eigenfunctions for various critical random matrix ensembles are investigated  in perturbation series in the regimes of strong and weak multifractality. In both regimes we obtain expressions similar to those of the critical banded random matrix ensemble extensively discussed in the literature. For certain ensembles, the leading-order term for weak multifractality can be calculated within standard perturbation theory. For other models such a direct approach requires modifications which are briefly discussed.  Our analytical formulas are in good agreement with numerical calculations.  
\end{abstract}

\maketitle

\section{Introduction}

One of the main results of quantum chaos is the statement that eigenvalues and eigenfunctions of different physical quantum systems in the semiclassical limit are well described  by universal ensembles of random matrices. When the corresponding classical model is integrable, the Berry-Tabor conjecture \cite{berry} says that, once properly rescaled by the mean level density, quantum eigenvalues  obey the Poisson statistics, i.e. they behave as eigenvalues of diagonal random matrices with  independent diagonal elements.   According to the Bohigas-Giannoni-Schmit conjecture \cite{bohigas}, quantum eigenenergies  of classically chaotic systems on the other hand are distributed as eigenvalues of the standard ensembles of random matrix theory (RMT), where all  matrix elements are independent random variables and the measure is invariant with respect to the conjugation  over orthogonal, unitary, or symplectic matrices, depending only on the symmetries of the system \cite{mehta}.  A similar dichotomy  exists between eigenfunctions. For classically integrable systems eigenfunctions are strongly localized on quantized tori. For classically chaotic systems eigenfunctions are fully extended and can be approximated by a superposition of elementary solutions with coefficients being independent random variables \cite{berryrw}.

Naturally, there exist systems which are neither chaotic nor integrable, and their random matrix description, if any, is not an easy problem. The tight-binding Anderson model in three dimensions with on-site disorder and nearest-neighbor coupling \cite{anderson} is one of the most profoundly investigated examples. When disorder is not too large, its eigenfunctions are either localized or extended, depending on whether the corresponding eigenenergy is smaller or larger than an energy $E_c>0$ called the mobility edge, which depends on the strength of the disorder.  Eigenstates counted from the center of the band  with $|E|> E_c$ are exponentially localized and spectral statistics is close to the Poisson statistics. When $|E|< E_c$  states are extended and the statistical properties of eigenvalues are well described by the usual RMT statistics. A new phenomenon appears when $|E|\approx E_c$, called the metal-insulator transition (MIT) point.  In  \cite{mit} it was demonstrated numerically that statistical properties  of eigenvalues and eigenfunctions around this critical point differ considerably  from the standard statistics mentioned above. In particular, critical eigenfunctions  are neither localized nor extended but have multifractal features, which manifest in a non-trivial scaling of mean eigenfunction moments $\langle \sum_{i=1}^L|\Psi_i|^{2q}\rangle$  with the system size $L$. For localized states, moments do not depend on $L$, while for extended states they scale as $L^{-(q-1)d}$, where $d$ is the dimension of the system. For critical states the scaling is $L^{-(q-1)D_q}$, where $D_q\in[0,d]$, called the generalized (or multifractal) dimensions, are non-trivial functions of $q$.

Numerical calculations for the Anderson model are limited by the fact that the system is three-dimensional. In order to investigate properties of critical states, simpler models have been proposed 
(see e.g.~\cite{levitov,seligman}) in the form of $N\times N$ random matrices with elements slowly decreasing away from the main diagonal, such as
\begin{equation}
M_{mn}=p_m\delta_{mn}+V(m-n)
\end{equation}
where diagonal elements $p_m$ are independent random variables and off-diagonal elements $V(m-n)$ decrease as the first  power of the distance from the diagonal 
\begin{equation}
V(m-n)\underset{|m-n|\gg 1}{\sim} \frac{g}{|m-n|}.
\end{equation}
The mostly investigated model of this type is the critical banded random matrix ensemble (CrBRME) of real symmetric ($\beta=1$) or complex Hermitian ($\beta=2$) matrices \cite{seligman}, whose elements $M_{mn}$ are independently distributed Gaussian variables with zero mean and variance given by
\begin{eqnarray}
\langle |M_{nn}|^2\rangle&=&\frac{1}{\beta},\nonumber\\
\langle |M_{mn}|^2 \rangle &=&\frac{1}{2}\left [1+\Big (\frac{m-n}{g}\Big )^2 \right ]^{-1},\,\,\, m\neq n.
\label{CrBRME}
\end{eqnarray}
The results obtained for this model are reviewed in \cite{evers}. For all values of the coupling constant $g$ its eigenfunctions have nontrivial multifractal properties. Though fractal dimensions are not yet accessible to full analytical calculations, the construction of perturbation series yields an analytical approach to them. Perturbation series expansions of the $D_q$ were obtained for small and large values of the coupling constant $g$ \cite{mirlin,evers}. For  $g\ll 1$ (''strong multifractality'' limit, that is, the limit where states are almost localized) and $q>1/2$, the leading  term of the expansion into powers of $g$ for eigenstates close to the center of the spectrum $E=0$ was found to be \cite{note}
\begin{equation}
\label{banded_small_g}
D_q\approx 4 g\frac{\Gamma(q-1/2)}{\sqrt{\pi}\,\Gamma(q)}\left \{ \begin{array}{lc} \frac{1}{\sqrt{2}},&\beta=1\\ 
\frac{\pi}{2\sqrt{2}}, &\beta=2. \end{array}\right.
\end{equation}
When $g\gg 1$ (''weak multifractality'' limit, where states are almost extended), the expansion of $D_q$ into inverse  powers of $g$ for eigenstates close to the center of the spectrum reads
\begin{equation}
D_q=1-q \frac{1}{2\pi \beta g}+\mathcal{O}\left(\frac{1}{g^2}\right).
\label{banded_large_g}
\end{equation} 
Here we consider a different class of critical random matrix models, introduced in \cite{bogomolny}. They are constructed from Lax matrices of classical integrable one-dimensional systems of $N$ interacting particles whose positions and momenta are random variables. Integrability of the underlying systems makes it possible to analytically calculate various spectral properties, such as their joint distribution of eigenvalues and (in certain cases) several spectral correlation functions. A detailed investigation of spectral properties of these models was presented in \cite{bogomolny_giraud}. The expressions obtained for spectral correlation functions have unusual features but their properties are characteristic of critical systems, therefore it might be expected that eigenfunctions display multifractality properties. The purpose of this paper is to investigate numerically fractal properties of several different critical random matrix ensembles, and to calculate analytically multifractal dimensions in perturbation series in the two regimes of strong and weak multifractality. Some of the results have been briefly mentioned in \cite{entropy}. 

Let us introduce the four random matrix ensembles that we consider in the present paper. In all models, $g$ is a free parameter (coupling constant) independent on $N$. The first three ensembles are related with Lax matrices of the rational, hyperbolic, and trigonometric Calogero-Moser models, respectively \cite{perelomov}. The rational Calogero-Moser ensemble CM$_r$ is defined by $N\times N$ Hermitian matrices of the form
\begin{equation}
M_{mn}=p_m\delta_{mn}+\mathrm{i} g\frac{1-\delta_{mn}}{m-n}
\label{modelC}
\end{equation}
($\delta_{mn}$ is the Kronecker delta). The hyperbolic Calogero-Moser ensemble CM$_h$ consists of matrices
\begin{equation}
M_{mn}=p_m\delta_{mn}+\mathrm{i} g \frac{\mu(1-\delta_{mn})}{N\sinh \big [\mu(m-n)/N\big ] }\ ,
\label{modelD}
\end{equation}
with $\mu$ a real parameter independent on $N$. The trigonometric Calogero-Moser ensemble CM$_t$ is a set of matrices of the form
\begin{equation}
M_{mn}=p_m\delta_{mn}+\mathrm{i} g\frac{\mu(1-\delta_{mn})}{N\sin \big  [\mu(m-n)/N\big ]}\ .
\label{modelB}
\end{equation}
In all these ensembles, $p_m$ are independent random variables with zero mean and unit variance \cite{bogomolny}.

The fourth ensemble that we consider is the Ruijsenaars-Schneider (RS) ensemble, defined as the ensemble of $N\times N$ unitary matrices of the form
\begin{equation}
M_{mn}=\frac{\mathrm{e}^{\mathrm{i}\Phi_m}}{N}\frac{1-\mathrm{e}^{2\pi\mathrm{ i} g }}{1-\mathrm{e}^{2\pi\mathrm{ i}(m-n+g)/N}},
\label{modelA}
\end{equation}
with $\Phi_m$ independent random phases uniformly distributed between $0$ and $2\pi$. It is related with the Ruijsenaars-Schneider model of $N$ classical particles \cite{rs} and with the quantum map corresponding to the quantization of an interval-exchange map \cite{giraud, Schmit}. 

Eigenfunctions and eigenvalues of matrices $M_{mn}$ are defined by the equation
\begin{equation}
\sum_{n=1}^N M_{mn}\Psi_{n}(\alpha)=\lambda_{\alpha} \Psi_{m}(\alpha)
\label{eq_M}
\end{equation}
(here and below we label eigenvalues and eigenfunctions by Greek letters). We assume that eigenfunctions are normalized, $\sum_{n=1}^N|\Psi_{n}(\alpha)|^2=1$. The main object of our investigation is the asymptotic behavior of the mean moments of eigenfunctions
\begin{equation}
I_q=\frac{1}{N}\sum_{j=1}^N P_q(j)\ ,
\label{I_q}
\end{equation}
where
\begin{equation}
P_q(j)=\frac{1}{\rho(E)}\Big \langle \sum_{\alpha=1}^N|\Psi_j(\alpha)|^{2q}\delta(E-\lambda_{\alpha}) \Big \rangle 
\label{moments}
\end{equation}
is the local mean $q$th moment of eigenvalues. Here  $\rho(E)$ is the total mean eigenvalue density
\begin{equation}
\rho(E)=\frac{1}{N}\Big \langle \sum_{\alpha=1}^N \delta(E-\lambda_{\alpha})\Big \rangle 
\end{equation}
and $\langle \ldots \rangle$ denotes the average over the random matrix ensemble under consideration. Multifractal exponents $D_q$ characterize the asymptotic behavior of the mean moments when the matrix dimension $N$ goes to infinity. They are defined by the scaling
\begin{equation}
I_q \underset{N\to \infty}{\sim} N^{-(q-1)D_q}.
\label{D_q}
\end{equation}
For localized states the mean moments do not depend on $N$ and thus $D_q=0$. For extended states $D_q$ equals the dimension of the system; in particular for matrix models eigenvectors are one-dimensional, and therefore $D_q=1$. Our purpose is to calculate the leading-order term of the perturbation series expansion of $D_q$. We consider two regimes, characterized by strong and weak multifractality of the eigenfunctions. In the strong multifractality case the unperturbed states are localized and the zeroth order of the multifractal dimension is $D_q^{(0)}=0$. In the weak multifractality case the unperturbed states are extended and $D_q^{(0)}=1$. When a perturbation is added, fractal dimensions are changed to $D_q=D_q^{(0)}+d_q$. The small correction $d_q$  is obtained by expanding eigenfunction moments $I_q$ in perturbation series: when $d_q \ln N\ll 1$,  one has
\begin{equation}
\label{Iqeps}
I_q\sim  N^{-(q-1)D_q^{(0)}}(1-(q-1)d_q \ln N).
\end{equation}

Perturbation series  at small values of the coupling constant corresponds to strong multifractality and its construction follows the same scheme as for the CrBRME ensemble \cite{levitov, mirlin, evers, ossipov}. Matrices of CM ensembles can be expressed as
\begin{equation}
M_{mn}=p_m\delta_{mn}+g(1-\delta_{mn})M_{mn}^{(1)},
\label{small_g_matrix}
\end{equation}
where diagonal elements $p_m$ are independent random variables distributed according to some probability density $\sigma(p)$. For matrices of the RS ensemble it is convenient to first rewrite the matrix (\ref{modelA}) as
\begin{equation}
M_{mn}=\delta_{mn}\frac{\mathrm{e}^{\mathrm{i}\Phi_m+\mathrm{i}\pi g(1-1/N)}}{N}\frac{\sin(\pi g)}{\sin(\pi g/N)}+(1-\delta_{mn})\frac{\mathrm{e}^{\mathrm{i}\Phi_m}}{N}\frac{1-\mathrm{e}^{2\pi\mathrm{ i} g }}{1-\mathrm{e}^{2\pi\mathrm{ i}(m-n+g)/N}},
\label{rewrite_A}
\end{equation}
When $g\to 0$ the limits of each term exists and after rescaling by a factor 
$\exp[-\mathrm{i}\pi g(1-1/N)]$ the matrix reduces at leading order to
\begin{equation}
\label{M1RS}
M_{mn}\simeq\mathrm{e}^{\mathrm{i}\Phi_m}\delta_{mn} -g(1-\delta_{mn})\frac{2\pi \mathrm{i}}{N}\mathrm{e}^{\mathrm{i}\Phi_m}\frac{1}{1-\mathrm{e}^{2\pi\mathrm{ i}(m-n)/N }}\ ,
\end{equation}
which is of the form of \eqref{small_g_matrix}. The expansions \eqref{small_g_matrix} and \eqref{M1RS} corresponds to the quasi-degenerate perturbation series, as different diagonal terms may be  close to each other. Following \cite{levitov, mirlin, evers, ossipov}, we obtain the first order of the multifractal dimension expansion into powers of $g$ by replacing the $N\times N$ matrix $M$ by all possible  $2\times 2$ submatrices of $M$ 
\begin{equation}
\left ( \begin{array}{cc} M_{mm} & M_{mn} \\ M_{nm} & M_{nn}\end{array} \right )
\label{2_times_2}
\end{equation}
and summing over all indices $m$ and $n$. The calculations in this regime 
are presented for our four models in Section~\ref{Strong_multifractality}. 

The situation in the regime of weak multifractality, where the unperturbed states are extended, is less simple. The CrBRME result \eqref{banded_large_g}  was derived by mapping the problem to the supermatrix sigma model. For our models (and other similar models as well) this approach seems not applicable and we use a more direct method. For the RS ensemble \eqref{modelA}, it is possible to construct a formal perturbation series expansion around any non-zero integer value of $g$. In Section~\ref{weak_multifractality_A}  we present detailed calculations of fractal dimensions for eigenvectors of this model at second order in $\epsilon$ for $g=1+\epsilon$ and in Appendix we briefly discuss the case $g=k+\epsilon$ with integer $k\geq 2$. Fractal dimensions are an asymptotic property of the matrix ensemble when $N\to\infty$, while perturbation series expansion requires to take $\epsilon\to 0$. It appears that in the case $g=k+\epsilon$ with $k\geq 2$, contrary to the case $k=1$, limits $\epsilon\to 0$ and $N\to\infty$ do not commute and terms proportional to $N$ appear in the expansion of eigenvalue moments \eqref{Iqeps}, which are expected to be canceled by contributions of higher order in $\epsilon$. We postpone detailed discussion of this cancellation mechanism to another publication \cite{future}. Here, in the case $k\geq 2$, we just extract the term proportional to $\ln N$ and neglect the contribution linear in $N$. The obtained results are in a good agreement with numerical simulations. 

For the Calogero-Moser ensembles \eqref{modelC}--\eqref{modelB}, the large $g$ limit is obtained by dividing \eqref{small_g_matrix} by $g$ and rescaling the variables, so that the models considered can be rewritten in the form 
\begin{equation}
M_{mn}=M_{mn}^{(0)}+\frac{1}{g}p_m\delta_{mn},
\label{rescaling_MM}
\end{equation}
where $M_{mn}^{(0)}$ is a simple matrix with, in general, non-degenerate spectrum. For large $g$ the second term can be considered as a perturbation and we use usual perturbation series formulas to calculate the fractal dimensions. The success of this approach depends to a large extent on the analytical accessibility of eigenvalues and eigenfunctions of the unperturbed matrix $M_{mn}^{(0)}$. In Section~\ref{weak_ multifractality_B_D} we calculate the asymptotic expansion of the fractal dimensions for large values of the coupling constant for eigenvectors of CM$_r$, CM$_h$, and CM$_t$ ensembles. It appears that for these ensembles, at second order in $1/g$, fractal dimensions are zero for states close to the center of the spectrum. Numerical results suggest that for all three CM ensembles fractal dimensions decrease exponentially with $g$ and therefore should be zero at any order of perturbation series in $1/g$.  By contrast, the perturbation series for  states close to the spectral ends diverges for large matrix dimensions.  Again, this  corresponds to the fact that limits $g\to \infty$ and $N\to\infty$ do not commute, and may be explained by  the localization of such edge states at large coupling constant \cite{future}. 

\section{Strong multifractality}\label{Strong_multifractality}
In this section we recall the procedure used in \cite{levitov, mirlin, evers, ossipov} and apply it to our systems in order to derive the perturbation series expansion of the multifractal dimensions for small values of the coupling constant $g$. We present the method for the Calogero-Moser ensembles \eqref{modelC}--\eqref{modelB}. Its adaptation to the  RS ensemble is straightforward. 

Let us consider a matrix ensemble of the form \eqref{small_g_matrix}. The first-order correction is obtained by considering only $2\times 2$ submatrices of $M$ in the sum (\ref{moments}). Setting $h=gM_{mn}^{(1)}$, these submatrices are of the form
\begin{equation}
\left ( \begin{array}{cc} p_m & h \\ h^* & p_n\end{array} \right ).
\label{ph2times2}
\end{equation}
Eigenvalues of this matrix  are given by 
\begin{equation}
\mu_{\pm}= \xi\pm  \sqrt{\eta^2+|h|^2}
\label{mu2times2}
\end{equation}
with $\xi=(p_m+p_n)/2$ and $\eta=(p_m-p_n)/2$. The corresponding eigenvectors are $(u_+,v_+)$ and $(u_-,v_-)$ with
\begin{equation}
u_{\pm}=\frac{h}{\sqrt{|h|^2+\Delta_{\mp}^2}},\,\, v_{\pm}=\frac{-\Delta_{\mp}}{\sqrt{|h|^2+\Delta_{\mp}^2}}
\label{uv2times2}
\end{equation}
and $\Delta_{\pm}=\eta \pm \sqrt{\eta^2+|h|^2}$. Taking into account only the $2\times 2$ submatrix \eqref{ph2times2} in (\ref{moments}) means that the sum runs only over the two eigenvalues $\mu_{\pm}$, so at leading order the correction to the contribution to $P_q(m)$ due to transitions to the state labelled by $n$ is
\begin{equation}
\label{contributionpm}
\frac{1}{\rho(E)}\Big \langle |u_+|^{2q}\delta(E-\mu_+)-\delta(E-\xi-\eta)+ |u_-|^{2q}\delta(E-\mu_-)\Big \rangle,
\end{equation}
where the term $\delta(E-\xi-\eta)$ corresponds to the zeroth order term. By assumption, $h$ is fixed, and ensemble average is obtained by averaging over random values of $p_m$ and $p_n$ with a probability distribution $\sigma(p)$. The expression \eqref{contributionpm} becomes
\begin{eqnarray}
& &\frac{2}{\rho(E)} \int \mathrm{d}\eta \mathrm{d} \xi \sigma(\xi+\eta)\sigma(\xi-\eta)
\left(\frac{|h|^{2q}}{(|h|^2+\Delta_{-}^2)^q}\delta(E-\mu_{+})-\delta(E-\xi-\eta)\right .\nonumber \\ 
&+&\left .
\frac{|h|^{2q}}{(|h|^2+\Delta_{+}^2)^q}\delta(E-\mu_{-})\right)
\label{average}
\end{eqnarray}
(the factor 2 comes from the Jacobian of the change of variables from $(p_m,p_n)$ to $(\xi,\eta)$).  The integral over $\xi$ is easily performed. Changing variables by setting $\eta=|h|\sinh t$ in \eqref{average}, one gets
\begin{eqnarray}
& &\frac{2|h|}{\rho(E)} \int\mathrm{d}t \cosh t\left(\frac{\sigma(E-|h|e^{-t})\sigma(E-|h|e^{t})}{(1+e^{-2t})^q}
-\sigma(E)\sigma(E-|h|(\mathrm{e}^{t}-\mathrm{e}^{-t}))\right . \nonumber\\
&+&\left .\frac{\sigma(E+|h|e^{t})\sigma(E+|h|e^{-t})}{(1+e^{2t})^q}\right)\ 
\label{firstinh}
\end{eqnarray}
for the leading correction to  $P_q(m)$. The distribution $\sigma$ coincides with the zeroth order of the level density $\rho$ since unperturbed matrices are diagonal with eigenvalues $p_j$. Thus at first  order in $h$ the functions of the form $\sigma(E\pm |h|\varphi(t))$ in \eqref{firstinh} can be replaced by $\rho(E)$ and the integration over $t$ can be extended over the whole axis. Equation \eqref{firstinh} reduces to
\begin{equation}
2|h|\rho(E) \int_{-\infty}^{\infty}\Big (\frac{2\cosh(qt)}{(2\cosh t)^{q}}-1\Big )\cosh (t) \mathrm{d} t=-2|h|\rho(E)\frac{\sqrt{\pi}\Gamma \Big (q-\frac{1}{2}\Big)}{\Gamma(q-1)}\ ,
\label{main}
\end{equation}
where we have used a continuation of the known integral \cite{bateman}
\begin{equation}
\int_0^{\infty}\frac{\cosh(2at)}{\cosh^{2b}(t)}\mathrm{d}x = 4^{b-1}\frac{\Gamma(b+a)\Gamma(b-a)}{\Gamma(2b)}.
\end{equation}
The mean moment $I_q$ defined by \eqref{I_q} is then obtained by summing this result over  $n\neq m$, recalling that $h=gM_{mn}^{(1)}$. Up to the first order in $g$ the mean moment is therefore given by
\begin{equation}
I_q=1- 2g \rho(E)\frac{\sqrt{\pi}\Gamma \Big (q-\frac{1}{2}\Big)}{\Gamma(q-1)}S
\label{first_order}
\end{equation}
with 
\begin{equation}
S=\Big \langle \frac{1}{N}\sum_{\overset{m,n=1}{m\neq n}}^N | M_{mn}^{(1)}| \Big \rangle.  
\label{Sfirst_order}
\end{equation}
The average is taken over remaining random variables entering $M_{mn}^{(1)}$ (if any). 
For all critical systems $S$ diverges logarithmically when $N\to\infty$, as
\begin{equation}
S=2 s\ln N +\mathcal{O}(1),
\label{S}
\end{equation}
where $s$ is a non-zero constant which depends on the model \cite{levitov}.  Identifying the coefficient of $\ln N$ in (\ref{Iqeps}) and (\ref{first_order}) one concludes that at first order in $g$ multifractal dimensions of eigenfunctions with energy close to $E$ are given by
\begin{equation}
D_q= 4 g \rho(E)\, s\, \frac{ \sqrt{\pi}\, \Gamma \Big (q-\frac{1}{2}\Big)}{ \Gamma(q)}.
\label{dqstrong}
\end{equation}
This expression is valid only in the region $q>1/2$.  In order to find an analytic expression for $q<1/2$ we rely on a symmetry that has been observed for the multifractal dimensions of many critical systems \cite{universality}. Namely, the anomalous exponents defined by $\Delta_q=(D_q-1)(q-1)$ are symmetric with respect to $q=1/2$. In other words, multifractal dimensions for $q<1/2$ are related to those for $q>1/2$ by $\Delta_q=\Delta_{1-q}$. This leads to the following analytic expression for $q<1/2$,
\begin{equation}
D_q= 1+\frac{q}{1-q}\left(4 g \rho(E)\, s\, \frac{ \sqrt{\pi}\, \Gamma \Big (\frac{1}{2}-q\Big)}{ \Gamma(1-q)}-1\right).
\label{dqstrong_sym}
\end{equation}

The explicit values of constant $s$ for different models can be calculated as follows. For the rational Calogero-Moser ensemble CM$_r$ one has $|M_{mn}^{(1)}|=1/|m-n|$. The sum over $m,n$ yields
\begin{equation}
\label{asymptCMr}
\frac1N\sum_{\overset{m,n=1}{m\neq n}}^N | M_{mn}^{(1)}|=2\sum_{k=1}^{N-1}\frac{1}{k}-2 
\underset{N\to\infty}{\sim} 2\ln N+\mathcal{O}(1),
\end{equation}
thus $s=1$. A similar sum appears in the CrBRME model \eqref{CrBRME}. For this ensemble one has
\begin{equation}
|M_{mn}^{(1)}|\simeq\frac{|z_{mn}|}{\sqrt{2}|m-n|}\ ,
\end{equation}
where $z_{mn}$ is a Gaussian random variable with zero mean and unit variance which is real for $\beta=1$ and complex for $\beta=2$. Therefore $ \langle|z|\rangle=\sqrt{2/\pi}$ for $\beta=1$ and $\sqrt{\pi}/2$ for $\beta=2$, and summation over $m,n$ yields, using \eqref{asymptCMr}, $s=1/\sqrt{\pi}$ for $\beta=1$ and $s=\sqrt{\pi/8}$ for $\beta=2$, which allows to recover \eqref{banded_small_g} with $\rho(E)\simeq \rho(0)=1/\sqrt{2\pi}$. For the hyperbolic Calogero-Moser ensemble CM$_h$, one has
\begin{equation}
\label{asymptCMh}
\frac1N\sum_{\overset{m,n=1}{m\neq n}}^N | M_{mn}^{(1)}|=\frac{2\mu}{N^2}\sum_{k=1}^{N-1}\frac{N-k}{
\sinh(\mu k/N)}\underset{N\to\infty}{\sim} 2\ln N+\mathcal{O}(1)
\end{equation}
and, as for CM$_r$, $s=1$. For the trigonometric Calogero-Moser ensemble CM$_t$ one has
\begin{equation}
\label{asymptCMt}
\frac1N\sum_{\overset{m,n=1}{m\neq n}}^N | M_{mn}^{(1)}|=\frac{2\mu}{N^2}\sum_{k=1}^{N-1}\frac{N-k}{
|\sin (\mu k/N)|}.
\end{equation}
The main contribution to the sum \eqref{asymptCMt} comes from terms with $k/N$ close to zeros of $|\sin(\mu x)|$. The asymptotic behavior of the sum depends on the number of zeros when $0\leq x \leq 1$. One obtains $s=[\mu/\pi]$, where $[.]$ denotes the integer part. For the RS ensemble (\ref{modelA}), $M_{mn}^{(1)}$ is given by \eqref{M1RS} and it is easy to check that $s=1$.

\begin{figure}[ht]
\begin{center}
\includegraphics[width=.95\linewidth]{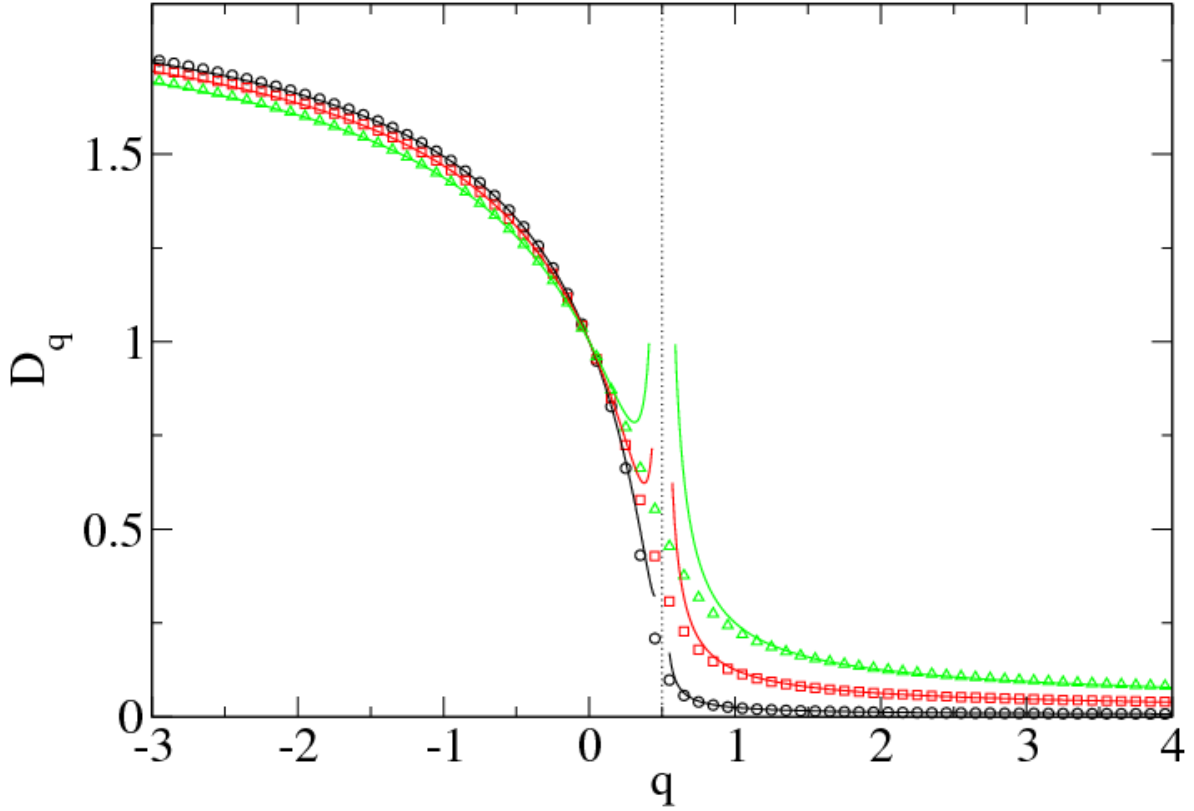}
\end{center}
\caption{(Color online) Fractal dimensions $D_q$ as a function of $q$ for CM$_r$ ensemble for $g=0.005$ (black circles), 0.025 (red squares) and 0.05 (green triangles). The $p_k$ are independent random variables distributed according to a Gaussian with mean 0 and variance 1. Symbols are numerical results (symbols are larger than error bars), solid lines correspond to the formulas \eqref{dqstrong} and \eqref{dqstrong_sym} with $\rho(E)=1/\sqrt{2\pi}$ and $s=1$. Matrix sizes for numerical fit are $N=2^n$, $8\leq n\leq 13$. Average is performed over the $N/16$ eigenvectors closest to the eigenvalue $E=0$. Number of random realizations of the matrix is between 2560 for $N=2^8$ and 40 for $2^{13}$. Dotted vertical line corresponds to $q=\frac12$.\label{strongInr}}
\end{figure}

\begin{figure}[ht]
\begin{center}
\includegraphics[width=.95\linewidth]{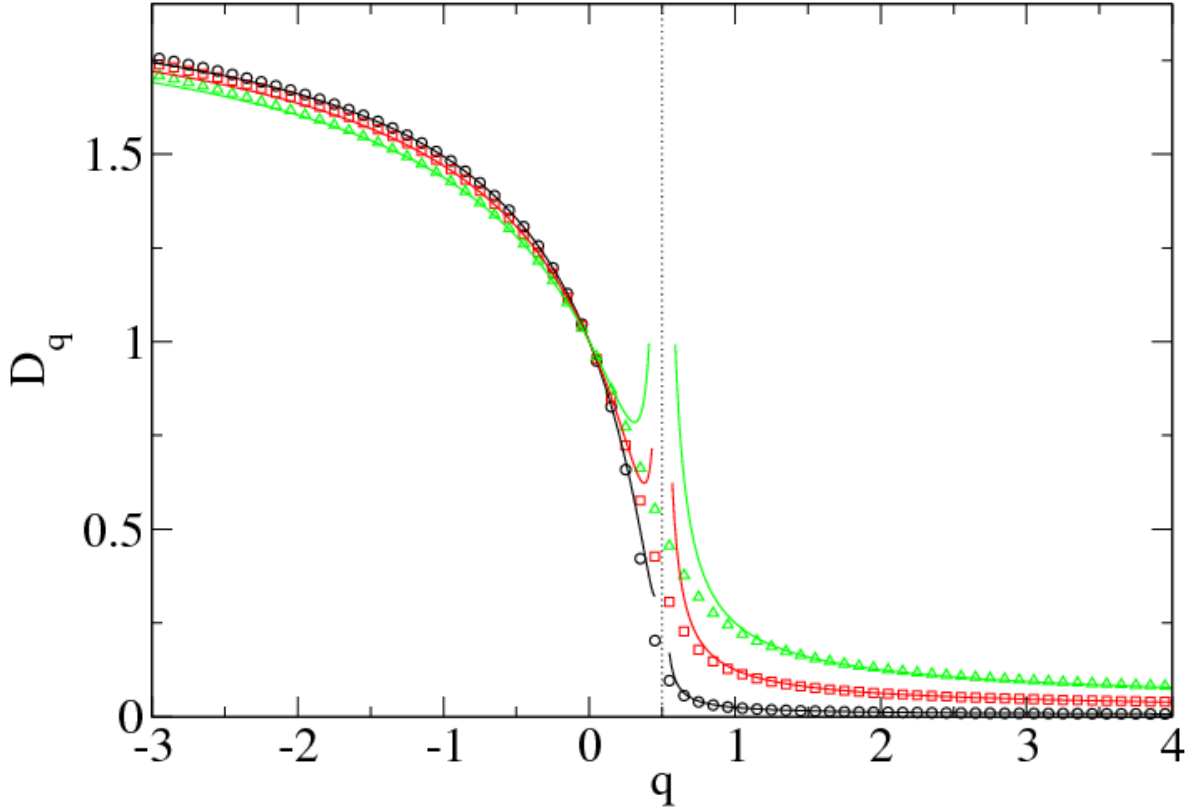}
\end{center}
\caption{(Color online) Same as Fig.~\ref{strongInr} for CM$_h$ ensemble with $\mu=2\pi$. The formula of Eqs.~\eqref{dqstrong} and \eqref{dqstrong_sym} is the same as for ensemble CM$_r$ ($s=1$).\label{strongIInr}}
\end{figure}

\begin{figure}[ht]
\begin{center}
\includegraphics[width=.95\linewidth]{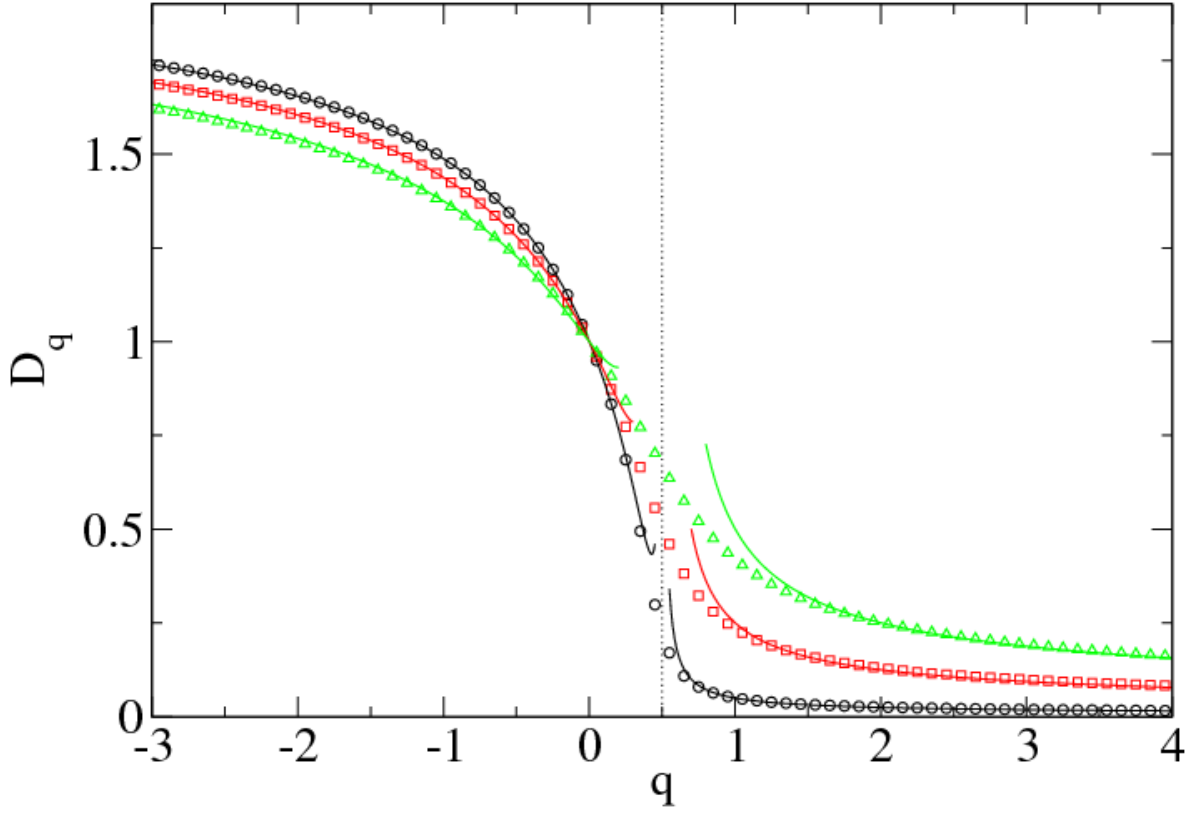}
\end{center}
\caption{(Color online) Same as Fig.~\ref{strongInr} for CM$_t$ ensemble with $\mu=2\pi$. Here matrix sizes are $N=2^n+1$, $8\leq n\leq 13$. The formula of Eqs.~\eqref{dqstrong} and \eqref{dqstrong_sym} is the same as for model CM$_r$ but with  $s=[\mu/\pi]=2$.\label{strongIIInr}}
\end{figure}

\begin{figure}[ht]
\begin{center}
\includegraphics[width=.95\linewidth]{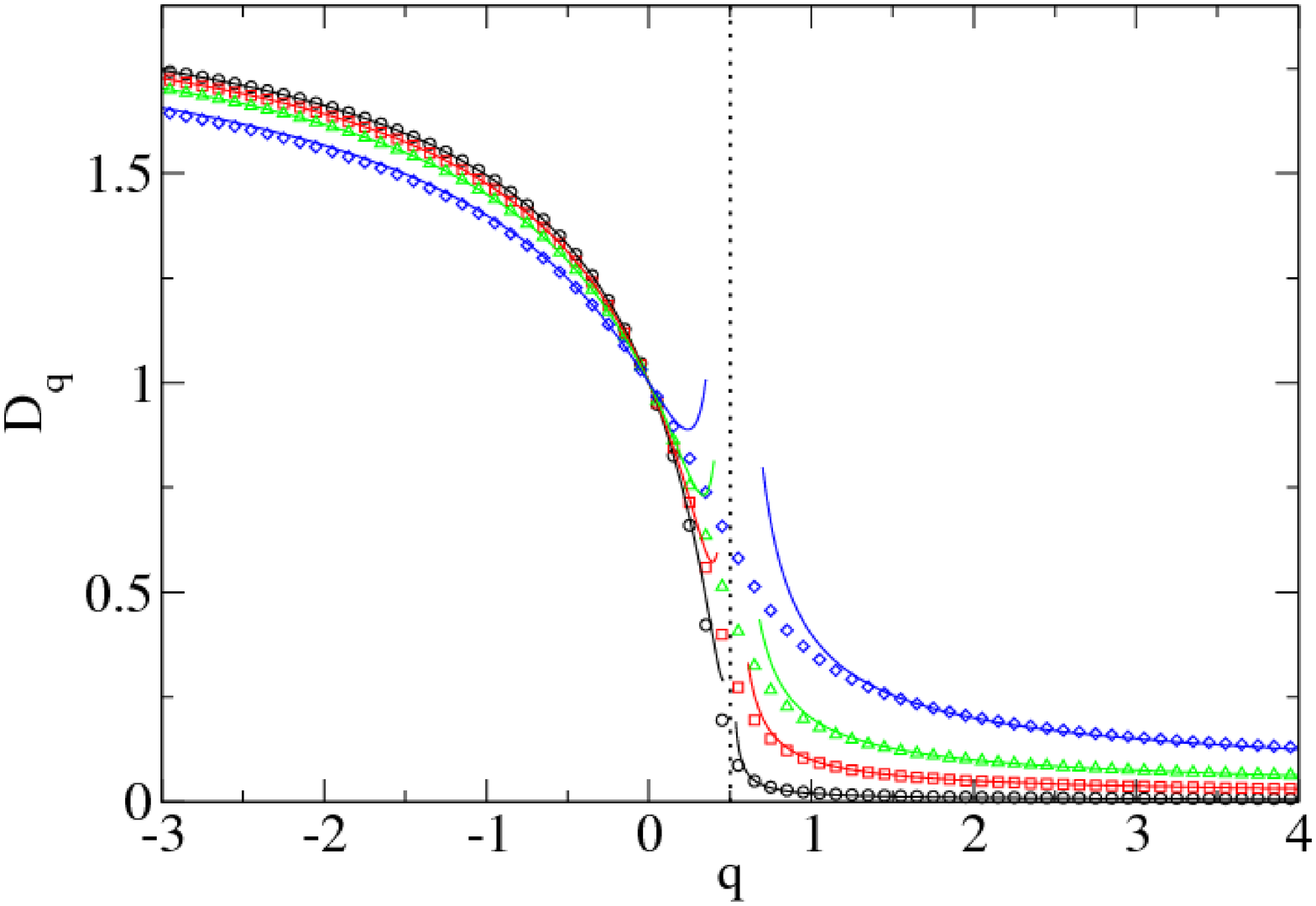}
\end{center}
\caption{(Color online) Fractal dimensions $D_q$ as function of $q$ for RS ensemble for $g=0.01$ (black circles), 0.05 (red squares), 0.1 (green triangles) and 0.2 (blue diamonds). The $\Phi_k$ are independent random variables distributed uniformly in $[0,2\pi]$. Symbols are numerical results (symbols are larger than error bars), solid lines correspond to the formula of Eqs.~\eqref{dqstrong} and \eqref{dqstrong_sym} with $\rho(E)=1/(2\pi)$ and $s=1$. Matrix sizes for numerical fit are $N=2^n$, $8\leq n\leq 12$. Average is performed over all eigenvectors. Number of random realizations of the matrix is from 32 for $N=2^8$ to 2 for $2^{12}$. Dotted vertical line corresponds to $q=\frac12$.\label{strongIIIb}}
\end{figure}

In order to assess the validity of the perturbation expansion formulas \eqref{dqstrong}--\eqref{dqstrong_sym} for strong multifractality, we performed detailed numerical calculations of multifractal dimensions for our four ensembles. Random realizations of matrices \eqref{modelC}--\eqref{modelA} are diagonalized, and a fit of the moments of the form $\log\langle \sum_i|\Psi_i|^{2q}\rangle=a+b\log N+c/N$ is obtained. For negative values of $q$, in order to avoid divergences due to exceedingly small values of the eigenfunction, a coarse-graining is first performed and the same fit as above is obtained for quantities $\log\langle \sum_i(\sum_j|\Psi_{4i+j}|^2)^q\rangle$. For CM ensembles, the average is performed over eigenvectors around the eigenvalue $E=0$, so that the density $\rho(E)$ is assumed to be constant over the range of vectors considered. For RS ensemble the average is performed over all eigenvectors. In Figs.~\ref{strongInr}--\ref{strongIIIb} we compare the numerical results for random matrix ensembles CM$_r$, CM$_h$, CM$_t$ and RS to formulas ~\eqref{dqstrong} and \eqref{dqstrong_sym}. Without any fitting parameter, the agreement is remarkable.

\section{Weak multifractality for RS ensemble}\label{weak_multifractality_A}
We now consider the opposite regime of weakly multifractal states. For RS ensemble (\ref{modelA}) this regime is reached when the coupling constant is $g=k+\epsilon$ with $\epsilon\ll 1$ and $k$ is any non-zero integer. Indeed, rewriting in this case matrix (\ref{modelA}) as in \eqref{rewrite_A} 
\begin{equation}
M_{mn}=\delta_{m+k,n}\frac{\mathrm{e}^{\mathrm{i}\Phi_m+\mathrm{i}\pi \epsilon(1-1/N)}}{N}\frac{\sin(\pi \epsilon)}{\sin(\pi \epsilon/N)}+(1-\delta_{m+k,n})\frac{\mathrm{e}^{\mathrm{i}\Phi_m}}{N}\frac{1-\mathrm{e}^{2\pi\mathrm{ i} \epsilon }}{1-\mathrm{e}^{2\pi\mathrm{ i}(m-n+k+\epsilon)/N}}
\label{rewrite_Ak}
\end{equation}
(here and below Kronecker symbols are to be understood modulo $N$), one sees that both terms have well-defined limits when $\epsilon\to 0$. The first term includes the constant phase factor $\mathrm{e}^{\mathrm{i}\pi \epsilon(1-1/N)}$. Rather than taking into account the expansion of this term, it is more convenient to redefine matrix (\ref{modelA}) to $\tilde{M}_{mn}=M_{nn}\mathrm{e}^{-\mathrm{i}\pi \epsilon(1-1/N)}$. Now one gets
\begin{equation}
\label{MM0M1}
\tilde{M}_{mn}=M_{mn}^{(0)}+\epsilon M_{mn}^{(1)} +\mathcal{O}(\epsilon^2)
\end{equation}
where 
\begin{equation}
M_{mn}^{(0)}=\mathrm{e}^{\mathrm{i}\Phi_m}\delta_{m+k,\,n }
\label{matrix_1}
\end{equation}
and 
\begin{equation}
M_{mn}^{(1)}=
-\frac{2\pi \mathrm{i}}{N}\mathrm{e}^{\mathrm{i}\Phi_m}\frac{1-\delta_{m+k,\,n}}{1-\mathrm{e}^{2\pi\mathrm{ i}(m+k-n)/N } }.
\label{matrix_k}
\end{equation}
In this section we consider the case $k=1$; the general case will be treated in the Appendix. For $k=1$, eigenfunctions $u_m(\alpha)$ and eigenvalues $\lambda_{\alpha}$ of the unperturbed matrix $M_{mn}^{(0)}$ obey the equation
\begin{equation}
\label{system}
\mathrm{e}^{\mathrm{i}\Phi_m}u_{m+1}(\alpha)=\lambda_{\alpha}u_{m}(\alpha)\ .
\end{equation}
Taking the product of both sides of this equation over all $m$ we see that eigenvalues can be chosen as 
\begin{equation}
\lambda_{\alpha}=\mathrm{e}^{\mathrm{i}\tilde{\Phi}+2\pi \mathrm{i} \alpha/N}\ ,
\label{theta_alpha}
\end{equation}
where $\tilde{\Phi}$  is the mean value of $\Phi_j$
\begin{equation}
\tilde{\Phi}=\frac{1}{N}\sum_{j=0}^{N-1}\Phi_j.
\end{equation}
Eigenfunctions of the unperturbed matrix are obtained recursively from (\ref{system}) by fixing one component, say $u_0(\alpha)=1/\sqrt{N}$, so that
\begin{equation}
\label{unalpha}
u_n(\alpha)=\frac{1}{\sqrt{N}}\mathrm{e}^{\mathrm{i}S_n(\alpha)}
\end{equation}
with
\begin{equation}
S_n(\alpha)=\frac{2\pi\alpha}{N} n -\sum_{j=0}^{n-1}\Phi_j+n\tilde{\Phi}\ .
\label{S_m}
\end{equation}
Contrary to the case $k=0$ where unperturbed eigenfunctions are localized, for $k\geq 1$ unperturbed eigenfunctions $u_n(\alpha)$ are extended.

When $g=1+\epsilon$, the perturbation of the matrix (\ref{modelA}) at first order in $\epsilon$ is given by \eqref{matrix_k} with $k=1$.  Let us expand exact eigenfunctions in a series of unperturbed ones
\begin{equation}
\Psi_n(\alpha)=u_n(\alpha)+\sum_{\beta}C_{\alpha \beta}u_n(\beta).
\label{Psi_n_alpha}
\end{equation}
Coefficients $C_{\alpha \beta}$ can be expanded into series of $\epsilon$. The expansion of $C_{\alpha \beta}$ with $\beta \neq \alpha$ starts with the first order in $\epsilon$ and $C_{\alpha \alpha}$ with the second order. The modulus square of the eigenvalue components is
\begin{eqnarray}
|\Psi_n(\alpha)|^2&=&|u_n(\alpha)|^2+\sum_{\beta}\Big [u_n^{*}(\alpha)C_{\alpha \beta}u_n(\beta)+u_n(\alpha)C_{\alpha \beta}^{*}u_n^{*}(\beta) \Big ]\nonumber\\
&+&\sum_{\beta, \gamma}C_{\alpha \beta}C_{\alpha \gamma}^{*}u_n(\beta) u_n^{*}(\gamma)=\frac{1}{N}(1+A_n(\alpha)+B_n(\alpha)),
\label{psi2bis}
\end{eqnarray}
where
\begin{equation}
A_n(\alpha)=\sum_{\beta}\Big [\mathrm{e}^{\mathrm{i}S_n(\beta)-\mathrm{i}S_n(\alpha)}C_{\alpha \beta}
+\mathrm{e}^{-\mathrm{i}S_n(\beta)+\mathrm{i}S_n(\alpha)} C_{\alpha \beta}^{*} \Big ]
\label{An}
\end{equation}
and
\begin{equation}
B_n(\alpha)=\sum_{\beta, \gamma}\mathrm{e}^{\mathrm{i}S_n(\beta)-\mathrm{i}S_n(\gamma)} C_{\alpha \beta}C_{\alpha \gamma}^{*}.
\end{equation}
Taking \eqref{psi2bis} to the $q$th power we get, up to the second order in $\epsilon$,
\begin{equation}
\label{psi2qtemp}
|\Psi_n(\alpha)|^{2q}=\frac{1}{N^q}\left[1+ q( A_n(\alpha)+ B_n(\alpha)) +\frac{q(q-1)}{2}A_n(\alpha)^2 \right].
\end{equation}
Normalization of the wavefunction implies that 
\begin{equation}
1=\sum_{n=1}^N|\Psi_n(\alpha)|^2=1+\frac{1}{N}\sum_{n=1}^N\left(A_n(\alpha)+ B_n(\alpha)\right)\ ,
\end{equation}
and therefore from \eqref{psi2qtemp} we get
\begin{equation}
\sum_{n=1}^N|\Psi_n(\alpha)|^{2q}=N^{1-q}\left[1+\frac{q(q-1)}{2N} \sum_{n=1}^N A_n(\alpha)^2\right]\ .
\label{second_term}
\end{equation}
At leading order in $\epsilon$, coefficients $C_{\alpha \beta}$ with $\beta\neq \alpha$ are given by
\begin{equation}
C_{\alpha \beta}=\epsilon \frac{V_{\alpha \beta}}{\lambda_{\alpha}-\lambda_{\beta}}
\label{C_alpha_beta_1}
\end{equation} 
with
\begin{equation}
V_{\alpha \beta}= \sum_{mn}u_m^{*}(\beta)M_{mn}^{(1)}u_n(\alpha).
\label{V_alpha_beta}
\end{equation}
Using \eqref{theta_alpha}--\eqref{S_m} we obtain
\begin{equation}
\lambda_{\alpha}-\lambda_{\beta}=\mathrm{e}^{\mathrm{i}\tilde{\Phi}+2\pi \mathrm{i}\alpha/N}-\mathrm{e}^{\mathrm{i}\tilde{\Phi}+2\pi \mathrm{i}\beta/N}=
2\mathrm{i}\sin \frac{\pi}{N}(\alpha-\beta)\mathrm{e}^{\mathrm{i}\tilde{\Phi}+\pi \mathrm{i}(\alpha+\beta)/N}
\end{equation}
and
\begin{equation}
V_{\alpha \beta}=
\frac{\pi}{N^2}
\sum_{\overset{\scriptstyle{ n,m}}{ \scriptstyle{ n\neq m+1}}} \frac{\mathrm{e}^{\mathrm{i}S_n(\alpha)-\mathrm{i}S_m(\beta)+\mathrm{i}\Phi_m-\mathrm{i}\pi(m+1-n)/N}}{\sin [\pi(m+1-n)/N]}\ .
\label{defV}
\end{equation}
Consequently, 
\begin{equation}
C_{\alpha \beta}=-\frac{\mathrm{i}\pi \epsilon}{2 N^2\sin [\pi(\alpha-\beta)/N]}\sum_{\overset{\scriptstyle{ n,m}}{ \scriptstyle{ n\neq m+1}}}\frac{\mathrm{e}^{\mathrm{i}F_{\alpha \beta}(m,n)+\mathrm{i}\chi_{mn}(\mathbf{\Phi})}}
{\sin [ \pi(m+1-n)/N]}\mathrm{e}^{-\pi\mathrm{i}(m+1-n)/N}\ ,
\label{C_alpha_beta}
\end{equation} 
where 
\begin{equation}
F_{\alpha \beta}(m,n)=\frac{2\pi}{N} \alpha(n-\frac{1}{2}) -\frac{2\pi}{N} \beta (m+\frac{1}{2})\ 
\label{Fab}
\end{equation}
and the function $\chi_{mn}(\mathbf{\Phi})$ contains the dependence on random variables $\Phi_j$,
\begin{equation}
\label{chimn}
\chi_{mn}(\mathbf{\Phi})=-\sum_{j=0}^{n-1}\Phi_j+\sum_{j=0}^{m}\Phi_j+(n-m-1)\tilde{\Phi}\ .
\end{equation}
According to (\ref{second_term}), the fractal dimensions are obtained by calculating
\begin{equation}
\label{Walpha}
W(\alpha)=\frac{1}{N}\sum_{n=1}^N A_n(\alpha)^2=\sum_{\beta , \gamma}R(\alpha,\beta,\gamma)
\end{equation}
where
\begin{eqnarray}
R(\alpha,\beta,\gamma)&=& \frac{1}{N}\sum_{n=1}^N 
\Big [\mathrm{e}^{\mathrm{i}S_n(\beta)-\mathrm{i}S_n(\alpha)}C_{\alpha \beta}
+\mathrm{e}^{-\mathrm{i}S_n(\beta)+\mathrm{i}S_n(\alpha)} C_{\alpha \beta}^{*} \Big ]\nonumber\\
&\times &
\Big [\mathrm{e}^{\mathrm{i}S_n(\gamma)-\mathrm{i}S_n(\alpha)}C_{\alpha \gamma}
+\mathrm{e}^{-\mathrm{i}S_n(\gamma)+\mathrm{i}S_n(\alpha)} C_{\alpha \gamma}^{*} \Big ].
\label{Rab}
\end{eqnarray} 
Using expression \eqref{S_m} for $S_n(\alpha)$ and summing over $n$ in (\ref{Rab}) one gets
\begin{equation}
R(\alpha,\beta,\gamma)=C_{\alpha \beta}(C_{\alpha \gamma}^*\delta_{\beta\gamma}+C_{\alpha\gamma}\delta_{2\alpha-\beta,\gamma})+\mathrm{c.c.} 
\label{R_beta_gamma}
\end{equation} 
where c.c. denotes the complex conjugate. Finally at leading order in $\epsilon$
\begin{equation}
W(\alpha)=\sum_{\beta\neq\alpha}\left(\left|C_{\alpha \beta}\right|^2+C_{\alpha \beta}C_{\alpha,\,2\alpha- \beta+\rho N}^{*} \right)+\mathrm{c.c.}
\label{almost_final}
\end{equation}
where  $\rho=0$ or $\pm 1$ accounts for the fact that the delta function in \eqref{R_beta_gamma} is modulo $N$. 

\begin{figure}[ht]
\begin{center}
\includegraphics[width=.95\linewidth]{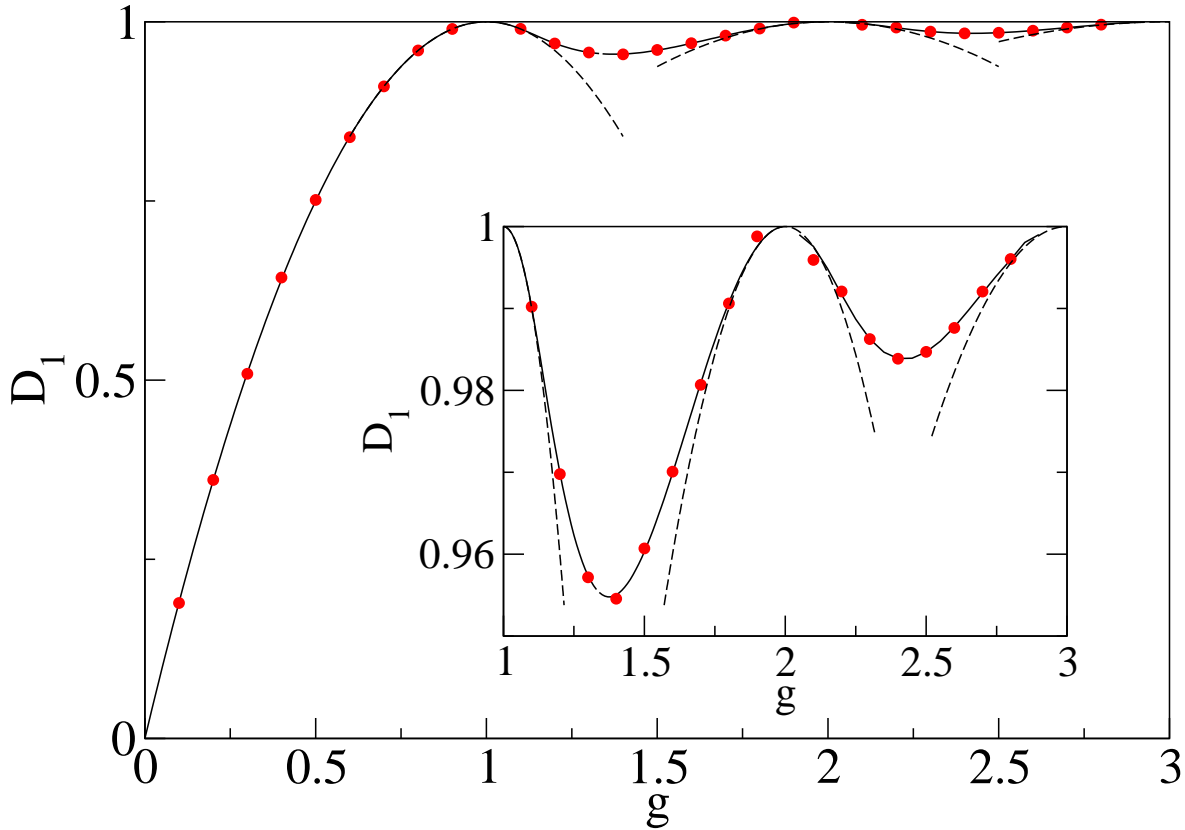}
\end{center}
\caption{(Color online) $D_1$ as function of $g$ for the RS model. Circles are numerical results
obtained by averaging over all eigenvectors taken from $128$ realizations for $N=2^8$ to $8$ realizations for $N=2^{12}$. Dashed lines indicate the leading perturbation series results (\ref{D_q_ruij}) for $q=1$. Solid line is an eye guide corresponding to the theoretical value given in \cite{bogomolny_giraud}, assuming that the conjecture of \cite{entropy} holds.
Inset: same data zoomed in for $1<g<3$. \label{fig_ruij}}
\end{figure}

In order to obtain fractal dimensions averaged over all eigenvectors of our matrix we need to calculate 
\begin{equation}
W=\frac{1}{N}\sum_{\alpha}W(\alpha).
\end{equation}
From (\ref{C_alpha_beta}) we have
\begin{eqnarray}
\label{C2}
|C_{\alpha \beta}|^2&=&\frac{\pi^2 \epsilon^2}
{4 N^4\sin^2[\pi(\alpha-\beta)/N]}\\
&\times&\hspace{-.5cm}\sum_{\overset{\scriptstyle{ n',m', n,m}}{ \scriptstyle{n'\neq m'+1, n\neq m+1}}}
\frac{\mathrm{e}^{-\mathrm{i}F_{\alpha \beta}(m',n')+\mathrm{i}F_{\alpha \beta}(m,n)-\mathrm{i}\chi_{m'n'}(\mathbf{\Phi})+\mathrm{i}\chi_{mn}(\mathbf{\Phi})}}
{\sin [ \pi (m'+1-n')/N]\sin [\pi(m+1-n)/N]}\mathrm{e}^{-\mathrm{i}\pi(m-m'-n+n')/N}\ .\nonumber
\end{eqnarray} 
The sum of $\left|C_{\alpha \beta}\right|^2$ over $\alpha$ and $\beta$ involves terms of the form
\begin{equation}
\sum_{\alpha}\sum_{\beta\neq\alpha}\frac{\mathrm{e}^{-\mathrm{i}F_{\alpha \beta}(m',n')+\mathrm{i}F_{\alpha \beta}(m,n)}}{\sin^2[\pi(\alpha-\beta)/N]}=
\sum_{\beta=1}^{N-1}
\frac{\mathrm{e}^{2\pi \mathrm{i}\beta(m'-m)/N}}{\sin^2[\pi\beta/N]}\sum_{\alpha}\mathrm{e}^ 
{2\pi \mathrm{i}\alpha(n-n'+m'-m)/N}
\end{equation}
which are non-zero if and only if $m-n=m'-n'$. We then take the average of \eqref{C2} over random phases $\Phi_j$ uniformly distributed in $[0,2\pi]$. One has to consider averaged quantities of the form
\begin{equation}
\label{moyennes}
\langle \mathrm{e}^{-\mathrm{i}\chi_{m'n'}(\mathbf{\Phi})}\mathrm{e}^{\mathrm{i}\chi_{mn}(\mathbf{\Phi})}\rangle ,
\end{equation}
which are non-zero if and only if coefficients of all the $\Phi_j$ in the exponential are zero. The condition $m-n=m'-n'$ cancels out terms in $\tilde{\Phi}$ in \eqref{moyennes}, and we have
\begin{equation}
\label{chiminuschi}
-\chi_{m'n'}(\mathbf{\Phi})+\chi_{mn}(\mathbf{\Phi})=\left\{
\begin{array}{ll}
\sum_{j=n}^{m}\Phi_j-\sum_{j=n'}^{m'}\Phi_j\,\,\, &m\geq n\\
-\sum_{j=m+1}^{n-1}\Phi_j+\sum_{j=m'+1}^{n'-1}\Phi_j\,\,\, &m\leq n-2\ .
\end{array}\right.
\end{equation}
Note that the case $m=n-1$ is excluded since it does not appear in the sum \eqref{C2}. In all cases, the only way to have all $\Phi_j$ vanish from \eqref{chiminuschi} is to have $m=m'$ and $n=n'$. Performing the remaining sum in \eqref{C2} under this condition, we get
\begin{equation}
\label{W1}
\langle\sum_{\alpha}\sum_{\beta\neq\alpha}|C_{\alpha \beta}|^2\rangle=
\frac{\pi^2 \epsilon^2 }{4 N^2}\sum_{\beta=1}^{N-1}\frac{1}{\sin^2(\pi\beta/N)}\sum_{n=1}^{N-1}\frac{1}{\sin^2(\pi n/N)}\ .
\end{equation}
The second term in \eqref{almost_final} can be calculated in the same way, using
\begin{equation}
C_{\alpha, 2\alpha-\beta+\rho N}=\frac{\mathrm{i}\pi\epsilon}{2 N^2\sin[\pi(\alpha-\beta)/N]}\sum_{\overset{\scriptstyle{ n,m}}{ \scriptstyle{ n\neq m+1}}}
\frac{\mathrm{e}^{2\pi \mathrm{i} [\alpha(n-2m-\frac12)+\beta(m-\frac12)]/N+\mathrm{i}\chi_{mn}(\mathbf{\Phi})}}{\sin [ \pi(m+1-n)/N]}\mathrm{e}^{-\pi\mathrm{i}(m+1-n)/N}\ .
\end{equation} 
The sum of $C_{\alpha \beta}^*C_{\alpha,\,2\alpha- \beta+\rho N}^{*}$ over $\alpha$ and $\beta$ involves terms of the form
\begin{equation}
\sum_{\alpha} \sum_{\beta\neq\alpha} \frac{\mathrm{e}^{-\mathrm{i}F_{\alpha \beta}(m',n')-2\pi\mathrm{i}[\alpha(n-2m-\frac12)+\beta(m-\frac12)]/N}}{\sin^2[\pi(\alpha-\beta)/N]}=
\sum_{\beta=1}^{N-1} \frac{\mathrm{e}^{2\pi \mathrm{i}\beta(m'-m)/N}}{\sin^2[\pi\beta/N]}
\sum_{\alpha} \mathrm{e}^{2\pi \mathrm{i}\alpha(n+n'-m'-m-2)/N}\ ,
\end{equation}
which are non-zero if and only if $n+n'= m'+m+2$. Again, when averaging over random phases, this latter condition cancels out terms in $\tilde{\Phi}$, and
\begin{equation}
\label{chiminuschi2}
-\chi_{m'n'}(\mathbf{\Phi})-\chi_{mn}(\mathbf{\Phi})=\left\{
\begin{array}{ll}
-\sum_{j=n}^{m}\Phi_j+\sum_{j=m'+1}^{n'-1}\Phi_j\,\,\, &m\geq n\\
-\sum_{j=n'}^{m'}\Phi_j+\sum_{j=m+1}^{n-1}\Phi_j\,\,\, &m\leq n-2\ .
\end{array}\right.
\end{equation}
All $\Phi_j$ terms vanish in the expressions above if and only if $m'=n-1$ and $n'=m+1$. Finally 
\begin{equation}
\label{W2}
\langle\sum_{\alpha}\sum_{\beta\neq\alpha}C_{\alpha \beta}^*C_{\alpha, 2\alpha-\beta+\rho N}^*\rangle=
-\frac{\pi^2\epsilon^2}{4 N^2}\sum_{\beta=1}^{N-1}\frac{1}{\sin^2(\pi\beta/N)}\sum_{n=1}^{N-1}\frac{\mathrm{e}^{-2\mathrm{i}\pi\beta n/N}}{\sin^2(\pi n/N)}\ .
\end{equation}
Inserting \eqref{W1} and \eqref{W2} into \eqref{almost_final} and performing the sum over $\beta$ using the following identity valid for integer $n$,
\begin{equation}
\sum_{\beta=1}^{N-1}
 \frac{\sin^2(\pi\beta n/N)}{\sin^2(\pi \beta/N)}=n(N-n),
\end{equation}
we obtain the final expression 
\begin{equation}
\langle W\rangle=\frac{\pi^2\epsilon^2}{N^3}\sum_{n=1}^{N-1}
 \frac{n(N-n)}{\sin^2(\pi n/N)}. 
\label{exact_second}
\end{equation}
 The summand in this expression diverges when $N\to\infty$  when $n$ is close to $0$ or $N$. One can rewrite the sum as 
\begin{equation}
\frac{\pi^2}{N^3}\sum_{n=1}^{N-1}
 \frac{n(N-n)}{\sin^2(\pi n/N)}=\frac{\pi^2}{N}\sum_{n=1}^{N-1}\left(\frac{n(N-n)}{N^2\sin^2(\pi n/N)}
-\frac{N}{\pi^2 n}-\frac{N}{\pi^2 (N-n)}\right)+2\sum_{n=1}^{N-1}\frac{1}{n}.
\end{equation}
For $N\to\infty$, the first sum tends to a finite value
\begin{equation}
\pi^2\int_{0}^{1}\left ( \frac{y(1-y)}{\sin^2(\pi y)}-\frac{1}{\pi^2 y}-\frac{1}{\pi^2 (1-y)}\right )\mathrm{d}y=2[1-\ln(2\pi)],
\end{equation}
while the second sum diverges as $\ln N$. Thus when $N\to\infty$ 
\begin{equation}
\langle W\rangle= 2 \epsilon^2 \ln N+\mathcal{O}(1).
\label{W}
\end{equation}
From (\ref{second_term}) and \eqref{Walpha} it follows that 
\begin{equation}
\frac{1}{N}\sum_\alpha\langle \sum_{n=1}^N|\Psi_n(\alpha)|^{2q}\rangle\underset{N\to \infty}{\sim} N^{1-q}\Big ( 1+q(q-1)\epsilon^2\ln N \Big )
\end{equation}
from which one extracts the multifractal exponents $D_q$ using \eqref{Iqeps} (recall that $g=1+\epsilon$)
\begin{equation}
D_q= 1-q(g-1)^2.
\label{Dqk1}
\end{equation}
In the Appendix this result is generalized to $g=k+\epsilon$ with integer $k\geq 2$. It is argued that fractal dimensions at leading order are then given by  
\begin{equation}
D_q= 1-q\frac{(g-k)^2}{k^2}.
\label{D_q_ruij}
\end{equation}
In Fig.~\ref{fig_ruij} we present the result of numerical calculations for $D_1$ in the RS model for different $g$. The agreement with the above perturbation series results is quite good. Numerical results suggest that \eqref{Dqk1} is actually the exact formula for the multifractal dimension $D_1$ of the RS ensemble in the whole range $0<g<1$. The calculation of higher-order terms of the perturbation series, though in principle possible, may be more complicated. The point is that in the limit $N\to\infty$ and fixed $\epsilon$ spectral properties of the RS ensemble are different for $\epsilon>0$ and $\epsilon<0$ \cite{bogomolny, bogomolny_giraud}. If the conjecture in \cite{entropy} is correct, it will signify that fractal dimensions for this ensemble also depend on the sign of $\epsilon$, and simple analytical results \eqref{Dqk1} and \eqref{D_q_ruij} may be valid only at second order in $\epsilon$.

\section{Weak multifractality for CM models}
\label{weak_ multifractality_B_D}
We now investigate fractal dimensions for the CM ensembles when the coupling constant $g$ goes to infinity. Matrices of these ensembles are defined as the sum of a diagonal term and a Hermitian off-diagonal matrix (see \eqref{modelC}--\eqref{modelB}). In order to apply the perturbation series analysis for large $g$ we redefine our matrices as $\tilde{M}_{mn}=M_{mn}/g$, so that 
\begin{equation}
\tilde{M}_{mn}=M_{mn}^{(0)}+\frac{1}{g}p_m\delta_{mn},
\label{rescaling_M}
\end{equation}
where $M_{mn}^{(0)}$ are the off-diagonal terms in \eqref{modelC}--\eqref{modelB} (divided by $g$). These matrices are non-random Toeplitz matrices, as their entries only depend on the difference $m-n$. The properties of such matrices when $N\to\infty$ are well investigated (see e.g.~\cite{basor} and references therein). 

Exact eigenvalues and eigenvectors of a Toeplitz  matrix  are not known in general. In the case of the CM$_t$ ensemble with $\mu=2\pi$ the matrix $M_{mn}^{(0)}$ becomes a circulant matrix, and as such its eigenvectors are simply given by $u_k(\alpha)=\exp(2\mathrm{i}\pi \alpha k/N)/\sqrt{N}$. The situation is thus similar to that of the previous Section (cf. \eqref{unalpha}), where unperturbed eigenvectors are extended. We now consider this case by closely following the approach of the RS ensemble. 

The unperturbed matrix reads
\begin{equation}
M_{mn}^{(0)}=\frac{2\pi\mathrm{i} (1-\delta_{mn})}{N\sin\left[2\pi(m-n)/N\right] }.
\end{equation} 
For simplicity we choose $N=2K+1$ as an odd integer. It is convenient to define eigenvectors in the following manner
\begin{equation}
\label{unal}
u_n(\alpha)=\frac{1}{\sqrt{N}}\exp \Big ( \frac{4\pi \mathrm{i}}{N}n \, \alpha \Big ) 
\end{equation}
with index $\alpha$ running from $-K$ to $K$. With such a choice, the corresponding eigenvalues take the simple form 
\begin{equation}
\label{lambda_unper}
\lambda_{\alpha}=-\frac{2\pi}{N}\sum_{n=1}^{N-1}\frac{\mathrm{i}}{\sin (2\pi n /N) }\mathrm{e}^{4\pi \mathrm{i}n \, \alpha/N }=-\frac{4\pi}{N}\alpha.
\end{equation} 
The spectrum is rigid, which implies that the density of unperturbed eigenvalues is constant in the interval $(-2\pi, 2\pi)$. All equations from \eqref{Psi_n_alpha} to \eqref{V_alpha_beta} remain valid, with $\epsilon=1/g$, $M_{mn}^{(1)}=p_m\delta_{mn}$ and $S_n(\alpha)=4\pi n \alpha/N$. From Eq.~\eqref{C_alpha_beta_1} one obtains
\begin{equation}
C_{\alpha \beta}=\frac{1}{g N(\lambda_\alpha-\lambda_\beta)}\sum_{m=1}^N p_m \mathrm{e}^{4\pi \mathrm{i}m \, (\alpha-\beta)/N }.
\label{C_alpha_beta_B}
\end{equation}
Equations \eqref{Walpha}--\eqref{almost_final} are unchanged, thus we just have to calculate $W(\alpha)$ from \eqref{almost_final} and \eqref{C_alpha_beta_B}. In both terms appearing in \eqref{almost_final}, the averaging over random variables $p_m$ involves expressions of the form
\begin{equation}
\sum_{m,m'=1}^N \langle p_mp_{m'} \rangle\mathrm{e}^{\pm 4\pi \mathrm{i}(m-m') (\alpha-\beta)/N }
=N\langle p^2\rangle.
\end{equation}
Thus after averaging over $p_m$ \eqref{almost_final} gives
\begin{equation}
\langle W(\alpha)\rangle=\frac{\langle p^2\rangle}{g^2 N}\sum_{\beta\neq\alpha}
\frac{2\lambda_\alpha-\lambda_\beta-\lambda_{2\alpha-\beta+\rho N}}{(\lambda_\alpha-\lambda_\beta)^2(\lambda_\alpha-\lambda_{2\alpha-\beta+\rho N})}\ .
\end{equation}
Since indices $\alpha$ and $\beta$ run between $-K$ and $K$, we have $2\alpha-\beta\in[-3K,3K]$. Replacing the unperturbed eigenvalues by their expression \eqref{lambda_unper} we get $2\lambda_\alpha-\lambda_\beta-\lambda_{2\alpha-\beta+\rho N}=4\pi\rho$, which means that for $2\alpha-\beta\in[-K,K]$ (i.e.~$\rho=0$) the contribution to $W$ vanishes. Consequently, contributions to (\ref{almost_final})  come only from terms with $\beta>2\alpha+K$ or $\beta<2\alpha-K$, yielding 
\begin{equation}
\langle W(\alpha)\rangle=\frac{N^2\langle p^2\rangle}{8\pi^2 g^2}T(\alpha,N) 
\end{equation}
where 
\begin{equation}
T(\alpha,N)=\sum_{j=K+1-|\alpha|}^{K+|\alpha|}
\frac{N^2}{j^2(N-j)}\, .
\end{equation}
This gives us the final expression for the averaged momenta of eigenvectors
\begin{equation}
\langle\sum_{n=1}^N|\Psi_n(\alpha)|^{2q}\rangle=N^{1-q}\Big [ 1+\frac{q(q-1)\langle p^2\rangle}{16\pi^2 g^2}T(\alpha,N)\Big ]. 
\label{psi_T}
\end{equation}
While the RS ensemble is an ensemble of unitary matrices, matrices corresponding to CM ensembles are Hermitian, and the position of the eigenvalue in the spectrum, specified by the value of the index $\alpha\in [-K,K]$, does matter. For eigenvectors whose eigenvalues are close to the center of the spectrum, the index is such that $|\alpha|\ll K$. For $|\alpha|$ fixed one has
\begin{equation}
T(\alpha,N) \underset{N\to\infty}{\longrightarrow}0,
\end{equation}
which, using \eqref{Iqeps}, implies that second-order correction to the multifractal dimension is 0 for these states. 
\begin{figure}[ht]
\begin{center}
\includegraphics[width=.7\linewidth]{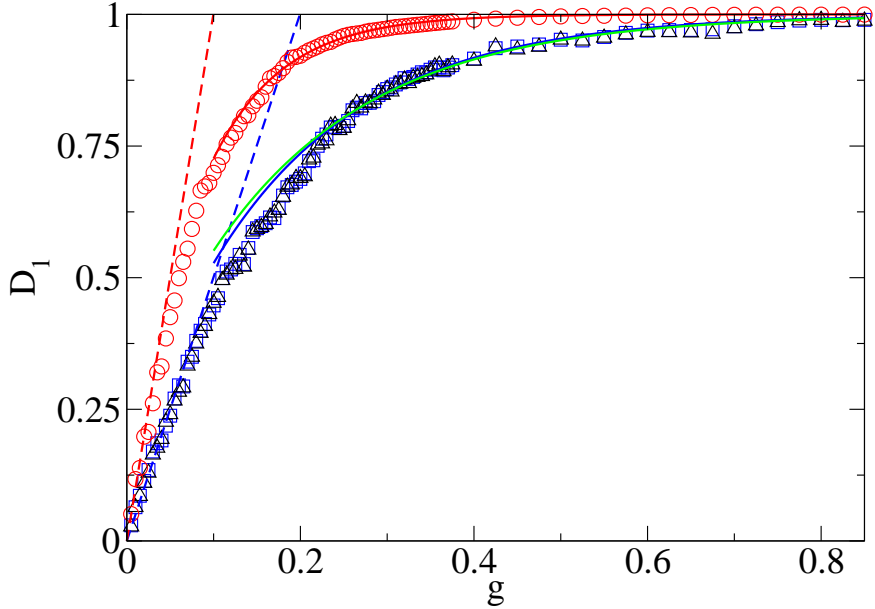}
\end{center}
\caption{(Color online) $D_1$ as a function of $g$ for models CM$_t$ (red circles), CM$_r$ (blue squares) and CM$_h$ (black triangles), obtained from 256 realizations of the random matrices for $N=2^8$ to 32 for $N=2^{11}$ (only $N/16$ central vectors are considered). Dashed lines indicate the leading perturbation series results for small $g$:  $D_1=2\sqrt{2\pi}s g$ with $s=1$ for models CM$_r$ and  CM$_h$, and $s=2$ for model CM$_t$. Solid lines indicate the fits for large $g$: $D_1\approx 1-0.85\mathrm{e}^{-5.8 g}$ for model CM$_r$, $D_1\approx 1-0.78\mathrm{e}^{-5.5 g}$ for model CM$_h$, and $D_1\approx 1-.92\mathrm{e}^{-12.2 g}$ for model CM$_t$.\label{fig_B_C}} 
\end{figure}

Numerical calculations, as presented in Fig.~\ref{fig_B_C}, suggest that in fact multifractal dimensions for Calogero-Moser models  at large $g$ are exponentially close to 1, for example, 
\begin{equation}
D_1\sim 1-\, C(g) \mathrm{e}^{-b g}
\label{exponential}
\end{equation}
with a certain slow varying function $C(g)$ and some parameter  $b$. If this is indeed the case, any order of perturbation series in $1/g$ would yield zero for states with $|\alpha|\ll K$.

By contrast, eigenfunctions corresponding to eigenvalues at the edge of the spectrum, labelled by $\alpha$ with $|\alpha|\simeq K$, are localized. Indeed, let $\nu=K-|\alpha|$ be the index of the eigenvector counted from the edge of the spectrum. One can show that for $\nu$ fixed $T(K-\nu,N)$ can be lower and upper bounded by functions which asymptotically behave as $N/(\nu+1)$ and $N/\nu$, respectively. The perturbation series expansion (\ref{psi_T}) is  valid formally only when the second term is much smaller than the first one, that is, only if $N\ll \xi$ where $\xi\simeq g^2 \nu$. Thus at fixed values of $g$ and $\nu$ \eqref{psi_T} is no longer a correct approximation for large $N$ and calculation of higher-order terms becomes important. In other words, limits $\epsilon\to 0$ and $N\to \infty$ do not commute. The breakdown of the above perturbation series seems to indicate that our assumption that  eigenfunctions with fixed $\nu$ are almost extended is not correct. Numerical simulations show that for large $g$ eigenvectors associated with eigenvalues located at the end of the spectrum are localized and that the localization length increases with the distance from the spectral boundaries. Rough estimation gives that the localization length is  proportional to $\xi=g^2\nu$. When $\xi\ll N$ states are strongly localized and fractal dimensions are zero. But when $\xi$ exceeds $N$ localization becomes  unessential and fractal dimensions take non-trivial values as in \eqref{exponential}. This situation is analogous to the 3-dimensional Anderson model where states close to spectral boundaries are localized. The existence or absence of a sharp mobility edge for the Calogero-Moser models requires further investigations. A more careful discussion of such boundary states in these and other models will be given elsewhere \cite{future}.

\begin{figure}[ht]
\begin{center}
\includegraphics[width=.95\linewidth]{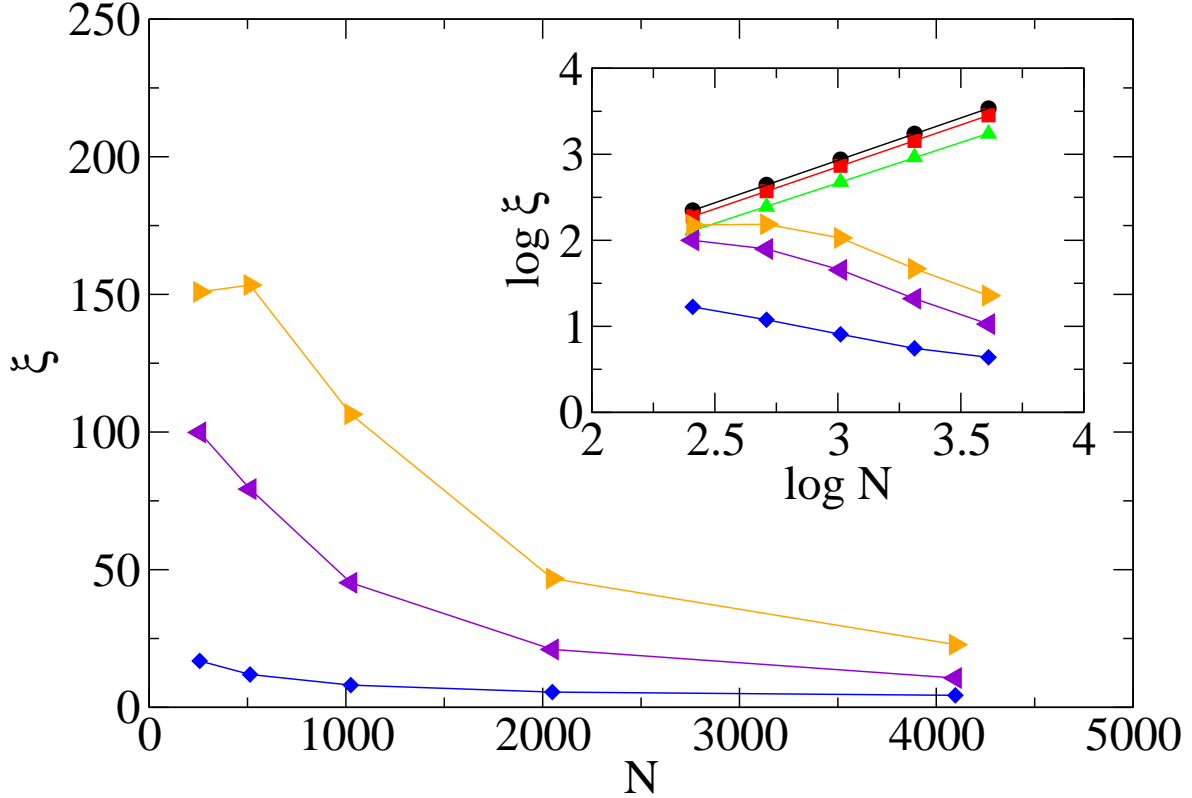}
\end{center}
\caption{(Color online) Inverse participation ratio $\xi=1/\langle\sum_i|\Psi_i(\alpha)|^4\rangle$ as a function of matrix size $N$ for model CM$_t$ for $g=0.5$ and $\mu=2\pi$. The average is taken over 32 realizations of the random matrix for $N=2^8$, to 2 realizations for $N=2^{12}$, and over eigenvectors associated with eigenvalues $\lambda_{2r k+1}<\ldots<\lambda_{2r(k+1)}$ with $r=8$ and $k=0,1,2$ from bottom to top (eigenvalues are ordered so that $\lambda_1<\ldots<\lambda_{N}$). We choose  $p_m$ as independent random variables distributed according to a Gaussian with mean 0 and variance 1. Inset: same as the main panel (three lower curves)  and eigenvectors averaged over $\lambda_{N/2-r},\ldots,\lambda_{N/2+r}$ (black circles), $\lambda_{N/4-r},\ldots,\lambda_{N/4+r}$ (red squares), $\lambda_{N/8-r},\ldots,\lambda_{N/8+r}$ (green triangles), $r=8$. For the three upper curves, solid lines are  linear fits with slopes respectively given by 0.99, 0.97, and 0.94. Logarithm is decimal.}
\label{ipr_IIInr}
\end{figure}

As an example, we display in Fig.~\ref{ipr_IIInr} the inverse participation ratio, which is the usual measure of the localization length, averaged over eigenvectors taken at different positions in the spectrum. It is clearly seen that states close to spectral boundaries are strongly localized while states far from boundaries are delocalized.

For models CM$_r$ and CM$_h$, where the unperturbed matrices are no longer circulant, there is (to our knowledge) no exact analytical expressions for the unperturbed eigenvectors and eigenvalues. Approximating eigenfunctions for these models as Fourier harmonics similar to 
\eqref{unal} and calculating eigenvalues as in \eqref{lambda_unper} one finds a good
agreement with the numerically computed spectrum. Using such an approximation we find that the behavior for large $g$ in all Calogero-Moser models is very similar.  Numerical results presented at Fig.~\ref{fig_B_C} agree with this conclusion. 

\section{Conclusion}
We investigated fractal dimensions of eigenfunctions for critical random matrix ensembles (\ref{modelC})--(\ref{modelA}) in perturbation series for strong and weak multifractality.  Spectral statistics of these ensembles are very different \cite{bogomolny} but fractal dimensions have many common points. 

For strong multifractality, when the coupling constant is small, fractal dimensions have the universal form
\begin{equation}
D_q= 4 g \rho(E)\, s\, \frac{ \sqrt{\pi}\, \Gamma \Big (q-\frac{1}{2}\Big)}{ \Gamma(q)},
\end{equation}
where $\rho(E)$ is the density of diagonal elements, and $s$ is a constant which depends on the system. For weak multifractality, fractal dimensions $D_q$ to leading order also have the universal form 
\begin{equation}
D_q=1-q t,
\end{equation}
with a constant $t$   depending on the system.  For the RS ensemble (\ref{modelA}) the weak multifractality regime corresponds to the vicinity of all integer points $g=k$ with $|k|\geq 1$, while for the CM ensembles it corresponds to $g\to \infty$.  The values of constants $s$ and $t$ for the models considered are  presented in Table~\ref{table1}. 

\begin{table}[ht]
\begin{tabular*}{0.8\textwidth}{@{\extracolsep{\fill}} || c || c | c | c | c | c | c ||}
\hline \hline
\hspace{.3cm} Model \hspace{.3cm}& \multicolumn{2}{ l}{ \hspace{.5cm} CrBRME \hspace{.4cm} } \vline & RS \hspace{.5cm} & CM$_t$ \hspace{.5cm} & CM$_r$ \hspace{.5cm} & CM$_h$ \hspace{.5cm} \\ 
 & $\beta=1$\hspace{.3cm} &   $\beta=2$ \hspace{.3cm} &  &  &  &  \\ \hline \hline 
$s$ & $1/\sqrt{\pi}$ \hspace{.3cm} &$\sqrt{\pi/8}$ \hspace{.3cm}  &   $1$ & $\left[\frac{\pi}{\mu}\right]$  & $1$ & $1$  \\ \hline \hline 
$t$ & $1/(2\pi g) \hspace{.3cm} $ &$1/(4\pi g)$\hspace{.3cm}  &\hspace{-0.3cm}  $(g-k)^2/k^2$\hspace{.3cm} & $0$  & $0$ & $0$    \\ \hline \hline 
\end{tabular*}
\caption{Constants $s$ and $t$ for different models.}
\label{table1}
\end{table}
For the CM ensembles the second order contribution to fractal dimensions for states far from spectral boundaries is zero. Numerical calculations suggest that fractal dimensions for these ensembles are exponentially small for large $g$ and, therefore, are zero in any finite order of perturbation series in $1/g$.   

The methods used in the paper are general and can, in principle,  be applied to a large variety of critical ensembles. Properties of strong multifractality  are well understood and controlled but weak multifractality limit for certain ensembles requires further investigation \cite{future}.    

\appendix

\section{Calculation of fractal dimensions in perturbation series for the RS ensemble in the general case}\label{app_k}

The purpose of this appendix is to calculate multifractal dimensions for the RS ensemble to leading order in $\epsilon$ when $g=k+\epsilon$ for integer $k\geq 2$.  The unperturbed matrix in this case is a matrix with shift by $k$,
\begin{equation}
\label{shift_k}
M_{mn}^{(0)}=\mathrm{e}^{\mathrm{i}\Phi_m}\delta_{m+k,n}\ .
\end{equation}
Eigenfunctions $u_m(\alpha)$ and eigenvalues $\lambda_{\alpha}$ obey the equation
\begin{equation}
\label{systemApp}
\mathrm{e}^{\mathrm{i}\Phi_m}u_{m+k}(\alpha)=\lambda_{\alpha}u_{m}(\alpha)\ .
\end{equation}
If one component  $u_{m}(\alpha)$ is fixed, then Eq.~\eqref{systemApp} allows to deduce all components of the form $u_{m+tk}(\alpha)$. Thus, when gcd$(k,N)$ is different from 1 one can construct families of solutions of the eigenvalue equation \eqref{systemApp} whose coefficients are nonzero only on subsets of indices of the form $t+rk$, $r=0,1,2,\ldots$. In what follows we denote $c=$gcd$(k,N)$ and $R=N/c$. Indices $n$ with $0\leq n\leq N-1$ are replaced by a pair of indices $(t,r)$ with $0\leq t\leq c-1$ and $0\leq r\leq R-1$ such that $n=t+r k$. Fixing $u_{t,0}(\alpha)=1/\sqrt{R}$, the other components are obtained recursively from \eqref{systemApp} as
\begin{equation}
\label{unalphaApp}
u_{t,r}(\alpha)=\frac{1}{\sqrt{R}}\mathrm{e}^{\mathrm{i}S_{t,r}(\alpha)}
\end{equation}
with
\begin{equation}
S_{t,r}(\alpha)=\frac{2\pi\alpha}{R} r -\sum_{j=0}^{r-1}\Phi_{t,j}+r\tilde{\Phi}_t\ .
\label{S_mApp}
\end{equation}
where $\tilde{\Phi}_t$  is the mean value of $\Phi_{t,j}$, defined by $\tilde{\Phi}_t=\frac{1}{R}\sum_{j=0}^{R-1}\Phi_{t,j}$. The corresponding eigenvalues can be chosen as 
\begin{equation}
\lambda_{t,\alpha}=\mathrm{e}^{\mathrm{i}\tilde{\Phi}_t+2\pi \mathrm{i} \alpha/R}\ .
\label{theta_alphaApp}
\end{equation}
The treatment is very similar to the case $k=1$, the only difference being that families of eigenvectors with fixed $t$ are treated in parallel. To simplify the discussion, let us now consider the case gcd$(k,N)=1$. The perturbation series of the matrix (\ref{modelA})  up to  the first order in $\epsilon$ is given by (\ref{MM0M1})--(\ref{matrix_k}). Equations \eqref{Psi_n_alpha}--\eqref{V_alpha_beta} are still valid and they yield
\begin{equation}
C_{\alpha \beta}=-\frac{\mathrm{i}\pi\epsilon}{2 N^2\sin [\pi(\alpha-\beta)/N]}\sum^{N-1}_{\overset{\scriptstyle{ n,m=0}}{ \scriptstyle{ n\neq m+1}}}\frac{\mathrm{e}^{\mathrm{i}F_{\alpha \beta}(m,n)+\mathrm{i}\chi_{mn}(\mathbf{\Phi})}}{\sin [ \pi(m+1-n)k/N]}\mathrm{e}^{-\pi\mathrm{i}(m+1-n)k/N}\ ,
\label{C_alpha_betaApp}
\end{equation} 
where $F_{\alpha \beta}(m,n)$ is given by \eqref{Fab}, and
\begin{equation}
\label{chimnApp}
\chi_{mn}(\mathbf{\Phi})=-\sum_{j=0}^{n-1}\Phi_{k,j}+\sum_{j=0}^{m}\Phi_{k,j}+(n-m-1)\tilde{\Phi}\ .
\end{equation}
A straightforward generalization of all calculations done for the $k=1$ case gives the final expression for $k\geq 1$
\begin{equation}
\langle W \rangle=\frac{\pi^2\epsilon^2}{N^3}\sum_{n=1}^{N-1}
\frac{n(N-n)}{\sin^2(\pi k n/N)}. 
\end{equation}
Terms divergent when $N\to\infty$ correspond to regions of $n$ close to zeros of the denominator, $n=mM+r$ where $M=[N/k]$, $m=0,1,\ldots k$, and $|r|\ll M$. Terms with $m=0$ and $m=k$ can be treated as in Section~\ref{weak_multifractality_A} and give a logarithmically divergent contribution $2\ln(N)/k^2$ but contributions of regions with $m=1,\ldots,k-1$ grow  linearly with $N$. Their calculation can be done by extending the summation over $r$ to all integers. Finally we gets  
\begin{equation}
\langle W\rangle = a N +\frac{2}{k^2}\ln N+\mathcal{O}(1)
\end{equation}
where
\begin{equation}
a=\frac{\pi^2}{k^4}\sum_{m=1}^{k-1} \frac{m(k-m)}{\sin^2(\pi ms/k)}
\end{equation}
and $s$ is the residue of $N$ modulo $k$ (we consider here the case of co-prime $N$ and $k$). Taking into account the logarithmic term results in \eqref{D_q_ruij}. 

The existence of linear in $N$ term, as above, indicates a breakdown of the simple perturbation series approach and the necessity of a more careful treatment  of higher-order terms.  The precise investigation of the compensation of such terms is beyond the scope of this paper and will be analyzed elsewhere \cite{future}.\\

\textit{Acknowledgements} -- EB is greatly indebted to  V.~Kravtsov,  I.~Lerner, A.~Ossipov, and O.~Yevtushenko for useful discussions.



\begin{thebibliography}{99}
\bibitem{note} For $\beta=1$ the expression we give here differs from the one in \cite{mirlin,evers} by a factor $\sqrt{2}$ due to the factor $1/2$ in our definition (\ref{CrBRME}) of the matrix ensemble.
\end{thebibliography}
\end{document}